\documentclass[submission,copyright,creativecommons]{eptcs}
\providecommand{\event}{QPL2020 Proceedings} 

\usepackage[utf8]{inputenc}
\usepackage{amsthm}
\usepackage{amsmath}
\usepackage{amsfonts}
\usepackage{mathtools}

\usepackage{graphicx}
\usepackage[export]{adjustbox}
\usepackage{floatrow}

\theoremstyle{plain}
\newtheorem{thm}{Theorem}[section]
\theoremstyle{definition}
\newtheorem{defn}[thm]{Definition}

\newtheorem{lema}[thm]{Lemma}

\providecommand{\urlalt}[2]{\href{#1}{#2}}
\providecommand{\doi}[1]{doi:\urlalt{http://dx.doi.org/#1}{#1}}

\begin{document}

\title{Contextuality and Expressivity of Non-locality}

\author{Sacha Huriot-Tattegrain \institute{ENS-Paris Saclay\\ France} \and  Mehdi Mhalla\institute{
Univ. Grenoble Alpes, CNRS, Grenoble INP, LIG,\\ F-38000 Grenoble France}}

\def\titlerunning{Contextuality and Expressivity of Non-locality}
\def\authorrunning{S.Huriot-Tattegrain \& M. Mhalla}
\maketitle
\begin{abstract}
    Contextuality is often referred to as a generalization of  non-locality. In this work, using  the hypergraph approach for contextuality we show how to associate a contextual scenario to a general $k$-partite non local game, and consider the reverse direction: how and when is it possible to represent a general contextuality scenario as a non local game. Using the notion of conditional  contextuality 
    we  show that it is possible to embed any contextual scenario in a two players non local game. We also discuss different equivalences of contextuality scenarios and show that the construction used in the proof is not  optimal by giving a simpler bipartite non local game when the contextual scenario is a graph    instead of a general hypergraph. 
    
\end{abstract}
\section{Introduction}
     Non-locality is a key feature for quantum information theory and has many implications on quantum computing.
     For example, there are certain linear algebra problems which can be solved by parallel quantum circuits of constant-depth but require logarithmic-sized classical circuits \cite{bravyi2018quantum}, or distributed computing tasks which can be solved in a constant number of rounds with a quantum setting but with a network-size number of rounds in the classical setting \cite{gall2018quantum}.\par
    These two approaches used the same quantum computing tool, non-local games. They are cooperation games in which the players are each asked a question and can't communicate, but they can share predefined quantum states and operate certain measurements on them according to their question which improves their success without sharing more information thanks to quantum superposition and entanglement. Contextuality is resource for models of quantum computation \cite{bermejo2017contextuality} which, for example, allows to describe the counter-intuitive correlations behind these strategies.\par
    It is usually acknowledged \cite{acin2015combinatorial} that non-locality is a particular case of contextuality, for instance Mermin Peres magic square game \cite{Mermin,Peres} was presented as a contextuality scenario. But, could certain aspects of non-locality capture contextuality as a whole? What is the exact relation between non-locality and contextuality? There are several ways to describe contextuality: with a hypergraphs-theoretic approach \cite{acin2015combinatorial}, a graph-theoretic one \cite{cabello2010non} and a sheaf-theoretic one \cite{abramsky2011sheaf}. 
    A partial result on this direction was obtained in the sheaf model in \cite{shane}, where it was proven that a large family of contextuality scenarios can be turned onto Bell scenarios. This paper uses the first approach and aims at obtaining any contextuality scenario from Foulis-Randall products. The players are identified with the factor scenarios of this product.\par
    
    First off we introduce the notion of contextuality as hypergraphs and several equivalence relations that are helpful to understand what need to be captured. Indeed, to show that it is possible to capture the properties of a hypergraph one may use a representative in an equivalence class. Then we express multipartite  non-locality with a contextuality formalism, we finally show in a constructive way how contextuality can be captured by bipartite non-locality, and show that the size of the construction can be optimized by showing a smaller embedding in the case where the contextuality scenario is a graph.  
\section{Hypergraph approach to Contextuality}
    We first define the graph theory tools which are going to be used to describe contextuality and equivalence relations. Many of these definitions and approaches are inspired from \cite{acin2015combinatorial}. The hypergraphs allow to see contextuality in the frame of general probabilistic theories \cite{barrett2007information}.
    \begin{defn}Contextuality Scenario \cite{acin2015combinatorial}\\
        A \textbf{contextuality scenario} is a hypergraph $H=(V,E)$ such that there is no isolated vertex, i.e.: \[V = \bigcup_{e\in E} e\] Its edges are called \textit{measurements} and its vertices are called the \textit{outcomes} (or \textit{results}) of the measurements containing them. Thereafter, the term \textit{hypergraph} will also be used to refer to a contextuality scenario and the \textit{size} of a scenario will refer to $|V|$.
    \end{defn}
    Within the interpretation of hypergraphs as experiments - that is in terms of preparation, settings and possible measurement - this definition ensures that each vertex is the result of at least one of the measurements.\par
    If a contextuality scenario isn't connected (if there are two sets of vertices with no edge in between them), its $m>1$ connected components can be interpreted as $m$ unrelated experiments, and the whole hypergraph as one parallel experiment.\par
    \begin{figure}[hb]
        \centering
        \includegraphics[width=7cm]{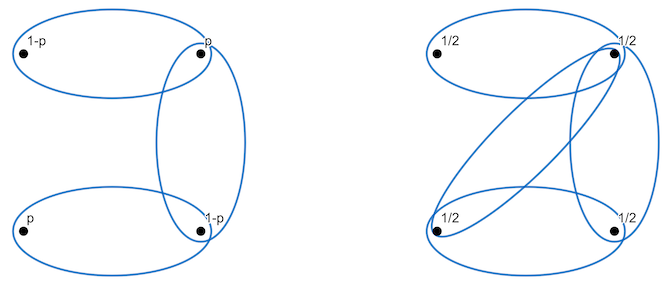}
        \caption{All models on the right scenario are of this form with $p\in[0,1]$ There is only one model for the left scenario}
        \label{fig:basic}
    \end{figure}
    Contextuality arises from considering the probability distributions over the outcomes:
    \begin{defn}General Probabilistic Model \cite{acin2015combinatorial}\\
        Let $H=(V,E)$ be a contextuality scenario, a \textbf{probabilistic model} on $H$ is a function $p:V\to [0,1]$ such that \[\forall e\in E,\ \sum_{v\in e} p(v) = 1\] The set of all probabilistic models on $H$ is denoted by $\mathcal{G}(H)$. And for any $W\subseteq V$, the sum $\sum_{v\in W} p(v)$ will be denoted by $p(W)$.
    \end{defn}
    Note that the likelihood to obtain a certain outcome is independent from which measurement containing it is conducted. Hence, when a measurement is conducted, one of its results is obtained with a probability given by this model.\par
    For any hypergraph $H$, different resources define different families of models: deterministic models $\mathcal{D}(H)$ are one were the outcome of each measurement is predefined (one vertex per edge), classical models $\mathcal{C}(H)$ is a convex combinations of deterministic models (deterministic hidden variables), and quantum models $\mathcal{Q}(H)$ are defined form measuring a quantum state on a Hilbert space (protective measurements). In \cite{acin2015combinatorial} it has been proved that $\mathcal{Q}(H)$ is characterized by a hierarchy of semi-definite programs that is enclosed into the local orthogonality models $\mathcal{CE}^1(H)$ (also called consistent exclusivity models) in which the sum of probability on any independent set of non-orthogonal vertices is smaller than 1 (where two vertices are called orthogonal if there exists an edge containing both of them).\par
    
    In order to say that a hypergraph captures another hypergraph: we define the notion of conditional contextuality to express that a subset of vertices of the first hypergraph can be identified with the vertices of the second in such a way that there is a bijection between the probability distributions  (the general probabilistic models) of the second and the probability distributions on the subset of vertices that can be extended to a general probabilistic model of the first. 
    \begin{defn}Conditional Contextuality\\
        Let $H=(V,E)$ and $H'=(V',E')$ be two hypergraphs such that there is an injection $\phi:V'\to V$. $H'$ is a \textbf{conditional contextuality scenario} of $H$ if and only if $\mathcal{G}(H)\circ\phi=\mathcal{G}(H')$, i.e.
        \begin{center}
            $\forall p\in\mathcal{G}(H),\ p\circ\phi\in\mathcal{G}(H')$\\
            $\forall p'\in\mathcal{G}(H'),\ \exists p\in\mathcal{G}(H),\ p\circ\phi=p'$
        \end{center}
    \end{defn}
    
    That is, the structure of a contextuality scenario $H'$ will be \textit{captured} by non-locality if one can find a non-local scenario $H$ whose models are extensions of all the models of the contextuality scenario $H'$. We will use this notion to try to capture the models of any contextuality scenario with on corresponding to a non local game (in the case of the two players the size will be of the form  $n_1 \times n_2$). \par

    In order to manipulate scenarios and reduce them while maintaining the core structure of their models, this following notion of sub-hypergraph will be useful:
    \begin{defn}Induced Sub-hypergraph \cite{acin2015combinatorial} \\
        Let $H=(V,E)$ be a hypergraph and $W\subseteq V$, the \textbf{sub-hypergraph} of $H$ induced by $W$ is the hypergraph \[H_W=(\ W\ ,\ \{\ e\cap W\ \colon\ e\in E\ \}\ )\] Any probabilistic model $p_W$ on $H_W$, can be extended to $p$ on $H$ by setting $p$ such that \[\forall v\in V,\  p(v)=\begin{cases}p_W(v) & \text{if }v\in W\\0 & \text{otherwise}\end{cases}\] In that case $p$ is a probabilistic model on $H$.\footnote{Note that  this extension is very different from the one of conditional contextuality as its models are just  the intersection of the probability models of the larger hypergraph  with the subspace where the missing vertices have 0 probability}
    \end{defn}
    
    \begin{figure}[ht]
        \centering
        \includegraphics[width=7cm]{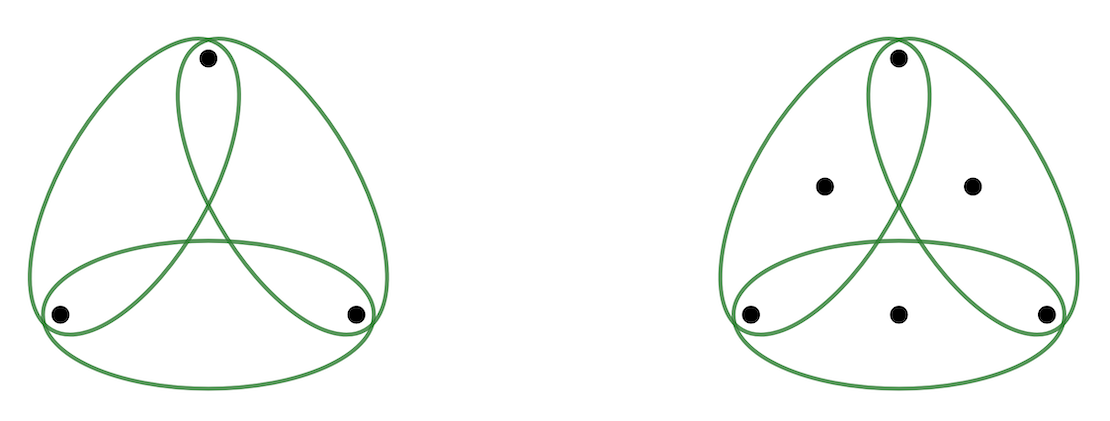}
        \caption{The contextuality scenarios $\Delta$ and $\Delta_3$, the first is an induced sub-hypergraph of the second}
        \label{fig:delta}
    \end{figure}

    In order to decrease the number of scenarios - and identify what is the fundamental structure of $\mathcal{G}(H)$ which makes $H$ more complex and hard to capture - these equivalence relations will aim to reduce the size of the hypergraphs considered. It will be also useful to try to define a canonical instance of contextuality scenarios.\par
    Because the majority of the following definitions rely on $\mathcal{G}(H)$, only scenarios such that $\mathcal{G}(H)\not=\emptyset$ will usually be considered for the examples and to illustrate the interpretation of these concepts.

    \begin{defn}Observational Equivalence\\
        Let $H=(V,E)$ and $H'=(V',E')$ be two contextuality scenarios. $H$ and $H'$ are \textbf{observationally equivalent} if and only if $\mathcal{G}(H)=\mathcal{G}(H')$. What is meant by this equation is than there exists a bijection $\phi:V'\to V$ such that \[\mathcal{G}(H')=\mathcal{G}(H)\circ\phi=\{\ p\circ\phi\ \colon\ p\in\mathcal{G}(H)\ \}\]
    \end{defn}
    The motivation behind decreasing the size of contextuality scenarios is that the constructions created to 'contain' are thus smaller. So, the three following definitions try to reduce $V$ and $E$ without losing the structure of the probabilistic models of the scenario.
    \begin{defn}Virtual Equivalence\\
        Let $H=(V,E)$ be a contextuality scenario and $e\subseteq V$. $e$ is a \textit{virtual edge} of $H$ if and only if \[\forall p\in\mathcal{G}(H),\ \sum_{v\in e} p(v) = 1\] $H$ is \textit{virtually included} in any $H'=(V,E\cup E')$ such that $E'$ contains only virtual edges of $H$. The symmetric closure of this inclusion is called \textbf{virtual equivalence}. The completion of $H$, denoted by $\overline{H}$, is the virtual equivalent of $H$ which has the most edges. 
    \end{defn}
    
    \begin{figure}[ht!]
        {\includegraphics[width=5cm]{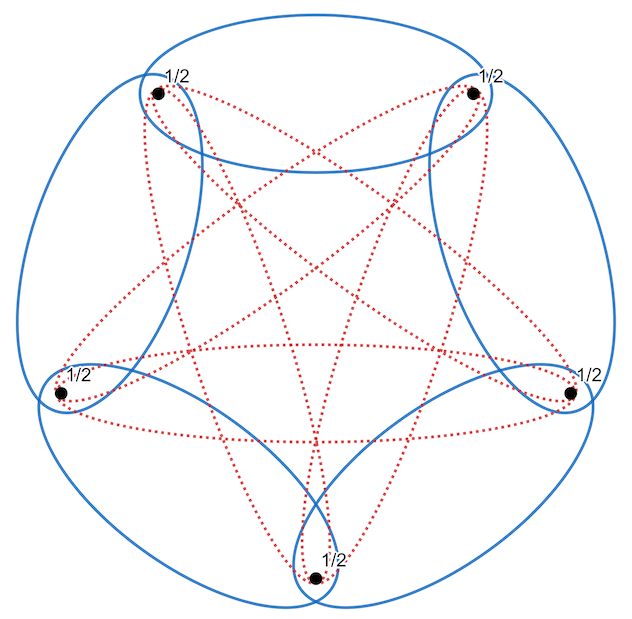}}
        {\caption{As the only model is $v\mapsto1/2$, any pair of vertices sum to $1$}
        \label{fig:complete}}
    \end{figure}

    \begin{defn} Equivalence by Contraction\\
        Let $H=(V,E)$ be a hypergraph and  $W$ a subset of vertices $W\subseteq V,\ W\not=\emptyset$. $W$ can be \textit{contracted} in $H$ if and only no edge $e$ cuts the subset into two non trivial parts:  \[\forall e\in E,\ (W\subseteq e)\vee(W\cap e=\emptyset)\] The vertices of $W$ are said to be \textit{indistinguishable}. For every $W'\subseteq W,\ W'\not=\emptyset$, the induced sub-hypergraph $H_{V\setminus(W\setminus W')}$ is a contraction of $H$. The symmetric transitive closure of this relation is called \textbf{equivalence by contraction}. Furthermore, there is an simple way to transform a probabilistic model $p$ on $H$ into a probabilistic model on $H_{V\setminus(W\setminus W')}$ as any function $p'$ such that \[\big(\forall v\in V\setminus W,\ p'(v)=p(v)\big)\ \wedge\ p'(W')=p(W)\] is a valid model of $H_{V\setminus(W\setminus W')}$.
    \end{defn}
   
    
        \begin{defn}Zero equivalence\\
        Let $H=(V,E)$ be a contextuality scenario and $W\subseteq V$. $H$ \textit{zero-reduces} to $H_W$ if and only if \[\forall p\in\mathcal{G}(H),\ \forall v\in V\setminus W,\ p(v)=0\] In that case, the vertices of $W$ are called \textit{zero weighted}. The symmetric closure of the zero-reduction is called \textbf{zero equivalence}.
    \end{defn}
    If there exists $e_1$ and $e_2$ in $E$ such that $e_1\subset e_2$, then $\forall p\in \mathcal{G}(H),\ \forall v\in e_2\setminus e_1,\ p(v)=0$. Only contextuality scenarios where there are no such edges are usually considered.\par
    \begin{figure}[ht]
        \centering
        \includegraphics[width=10cm]{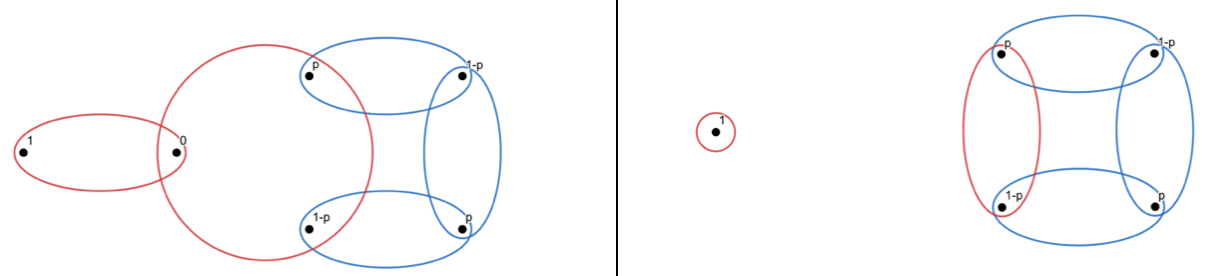}
        \caption{In any  model, the middle vertex has  weight  $0$}
        \label{fig:zero}
    \end{figure}
    The combined relation of these three equivalence relations is what we will consider in order to reduce the size of the studied scenarios. The non-local interpretation of a given contextuality scenario increases with the number of vertices and edges so this equivalence tries to build a smaller scenario with the same structure. 
    \begin{defn}VCZ Equivalence\\
        The \textbf{VCZ equivalence}, denoted $\sim_{vcz}$, is defined as the smallest equivalence relation containing virtual equivalence, equivalence by contraction and zero equivalence. For each scenario $H$, let $H^{vcz}=(V^{vcz},E^{vcz})$ denote an equivalent of $H$ with respect to $\sim_{vcz}$, obtained by removing all zero weighted vertices, by contracting the remaining vertices so that $|V^{vcz}|$ is minimal and finally by considering the completion and removing edges so that $|E^{vcz}|$ is minimal.
    \end{defn}
    $H^{vcz}$ is not unique but this notation can be used regardless because of the following result.
    \begin{thm} Consistency of VCZ Equivalence \\ \label{thm:vcz}
        Let $H_1$ and $H_2$ be two contextuality scenarios such that $H_1\sim_{vcz}H_2$, then (any choice of) $H_1^{vcz}$ and $H_2^{vcz}$ are observationally equivalent.
    \end{thm}
    Proof in  the Appendix.
\section{Expressing multipartite non local games as contextuality scenarios}

 A $k$ player combinatorial multipartite game is a game in which $k$ players not allowed to communicate receive each one local question $x_i $ in some set $X_i$  and provides an answer $a_i$ in some set $A_i$. The game is characterised by a winning global relation between the inputs  and outputs $W( x_1\ldots x_k,a_1\ldots a_k)$. 
 
 For each  player  $p_i$ each question $x_i$  is represented by an edge containing $|A_i|$ vertices.
 Each input correspond  to a player's measurement and each output to a possible outcome.
 
The scenario for one player is therefore a set of disjoint vertices that can be labeled $x|a$. We denote by $\mathcal{B}$ the set of such uncorrelated scenarios.

Combining scenarios can be done with Foulis Randall product.

  \begin{defn}Bipartite Foulis-Randall Product\\
        Let $H_A=(V_A,E_A)$ and $H_B=(V_B,E_B)$ be two hypergraphs. The sets of \textit{joint measurement} edges $E_{A\to B}$ et $E_{A\gets B}$, respectively of $H_A$ to $H_B$ and of $H_A$ from $H_B$, are defined as:
        \begin{align*}
            E_{A\to B}&:=\Bigg\{\bigcup_{a\in e_A}\big(\{a\}\times f(a)\big)\ \colon\ e_A\in E_A,\ f:e_A\to E_B\Bigg\}\\
            E_{A\gets B}&:=\Bigg\{\bigcup_{b\in e_B}\big(g(b)\times \{b\}\big)\ \colon\ e_B\in E_B,\ g:e_B\to E_A\Bigg\}
        \end{align*}
        The \textbf{bipartite Foulis-Randall product} of $H_A$ and $H_B$ is the product hypergraph $H_A\otimes H_B=(V_A\times V_B\ ,\ E_{A\to B}\cup E_{A\gets B})$.
    \end{defn}
    The set $E_{A\to B}$ (and symmetrically $E_{A\gets B}$) can be interpreted as Alice picking a measurement $e_A\in E_A$ and for each of its outcomes Bob is picking one measurement in $E_B$.\par

The set of global questions is  covered with the edges in  $E_{A\to B}\cap  E_{A\gets B}$.

As observed in \cite{acin2015combinatorial} by definition of the probability models  for any $p$,  for any two edges $\sum_{v\in e_1\setminus e_2 } p(v)=\sum_{v\in e_2\setminus e_1 } p(v)$. This allows to recover the nonsignaling condition  ensuring the setting in which the players do not communicate from the Foulis Randall product.

The bipartite game scenario is therefor just obtained by restricting the product viewing the winning relation as a rule:
 
\begin{defn} Bipartite Game\\
        Let $H=(V,E)= B_1\otimes B_2$ with $B_i=(V_i,E_i)\in\mathcal{B}$.\\
        The edges of $H$ of the form of $e_1 \times e_2  $ with $e_i\in E_i,\ \forall i\in[1,2]$ are called \textit{questions} of $H$ and the set of questions is denoted by $Q_E$.
        A function $r:Q_E\to\mathcal{P}(V)$ such that $\forall e\in Q_E,\ r(e)\subseteq e$ is called a \textit{rule} on $H$. For any rule $r$, the \textit{winning outcomes} are the vertices in $W_r=\bigcup_{e\in Q_E}r(e)$ and the \textbf{game on $H$ under the rule $r$} is the subsequent hypergraph \[H_r=\Bigg(W_r\ ,\ \{\ e\cap W_r\ \colon\ e\in E\ \}\Bigg)=H_{W_r}\]
    \end{defn}

\begin{figure}[ht!]
\centering
\begin{minipage}{.5\textwidth}
  \centering
  \includegraphics[width=5.2cm]{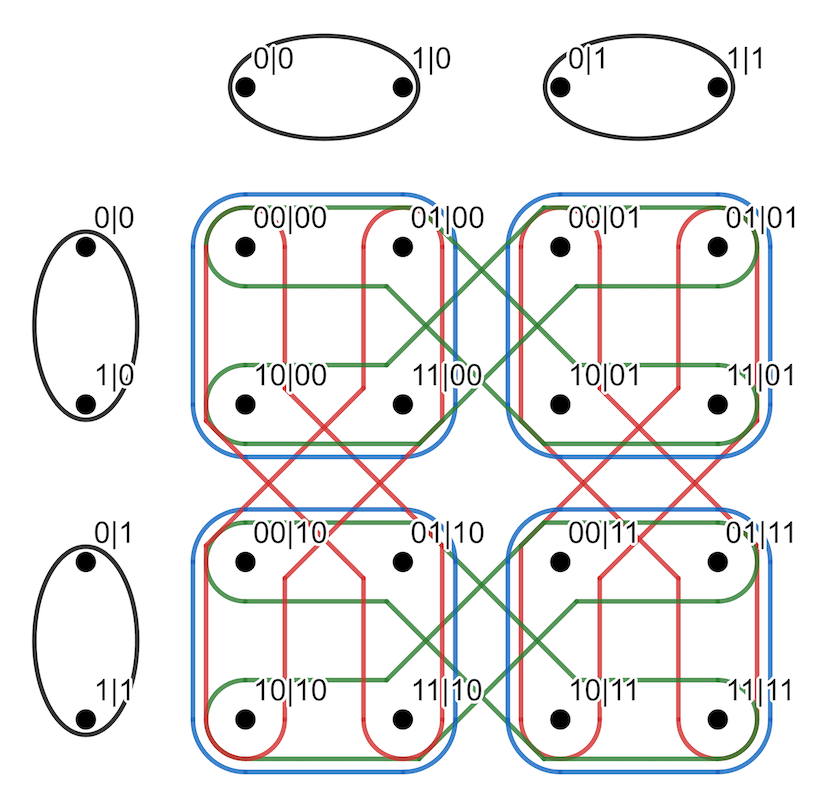}
  \label{fig:test1}
\end{minipage}%
\begin{minipage}{.5\textwidth}
  \centering
  \includegraphics[width=5.2cm]{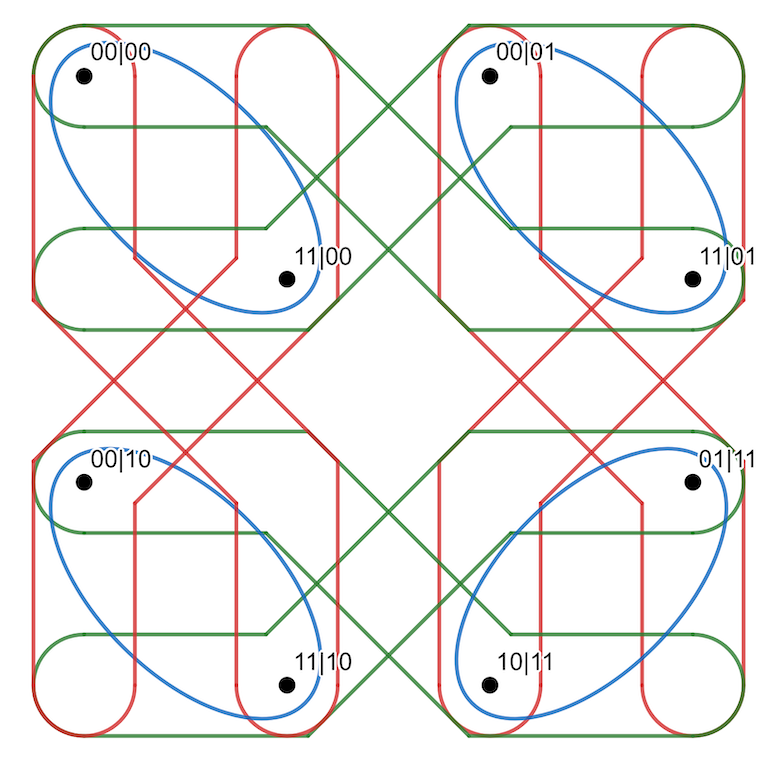}
  \caption{Left figure: Bell Scenario for 2 players with 2 measurements each which can give 2 outcomes, denoted as $B_{2,2,2}$. Right figure:  The CHSH game on the Bell scenario $B_{2,2,2}$, the only model is $v\mapsto1/2$ so the $\Delta$ scenario has a $2$-partite conditional interpretation}
  \label{fig:test2}
\end{minipage}
\end{figure}
     \newpage
    \begin{figure}[ht]
        \centering
        \includegraphics[width=10cm]{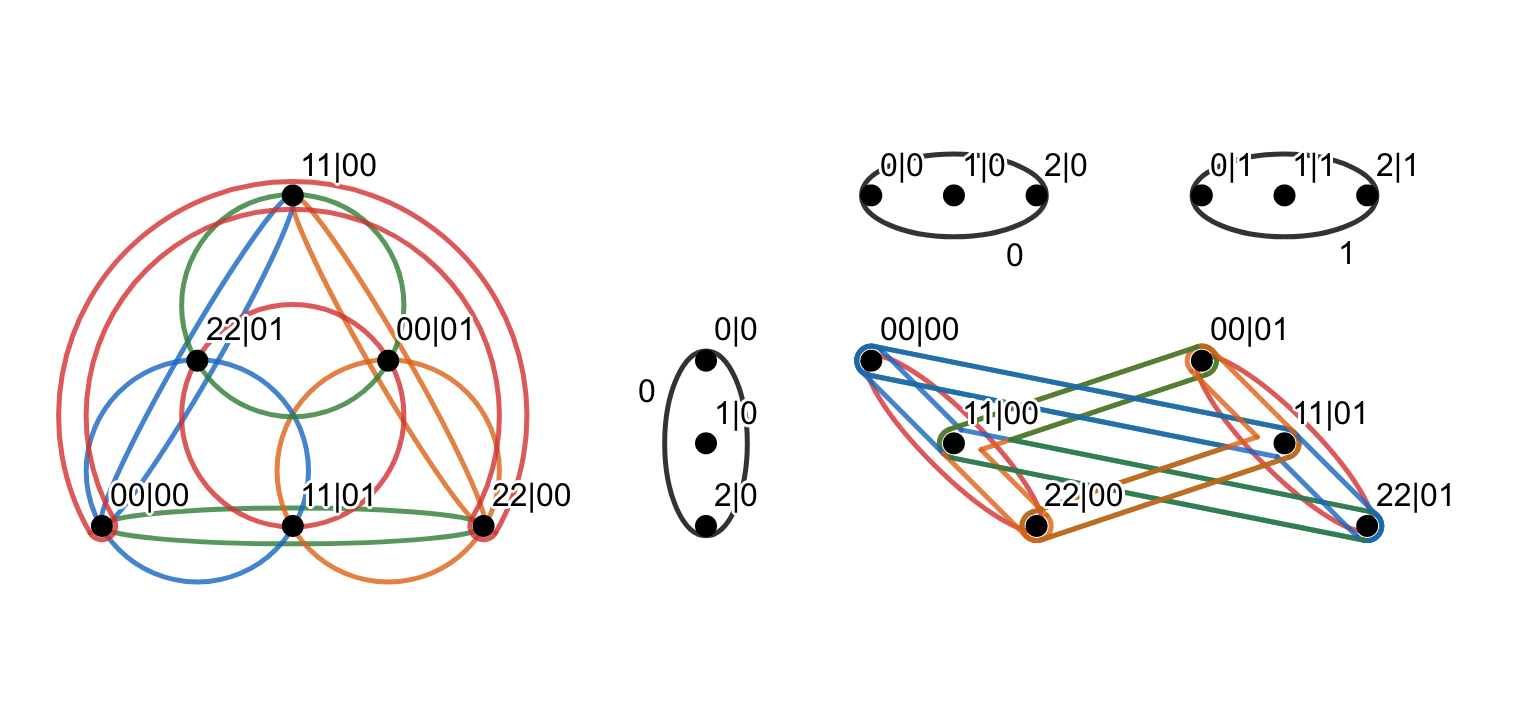}
        \caption{This contextuality scenario on the left can be directly obtained as a game with a 
         rule (denoted in red) which restrict the questions $00$ and $01$ to the diagonal answers $00$, $11$ and $22$}
        \label{fig:perfect}
    \end{figure}
     
     There are different choices of products of $k>2$ scenarios. However, all these choices are observationally equivalent- their completion does not depend on the particular choice of the product\cite{acin2015combinatorial}.

    \begin{defn} $k$-partite Joint Measurement\\
        Let $H_i=(V_i,E_i),\ i\in[1,k]$, be $k$ hypergraphs. The set of \textbf{joint measurement} edges $E_{i\to[1,k]\setminus i}$ of $H_i$ to the $H_{j\not=i}$, is defined as: \[E_{i\to[1,k]\setminus i}:= \left\{\Bigg\{(v_1,...,v_k)\ \colon\ \begin{array}{l} \forall j\not=i,\ v_j\in e_j \\ v_i\in f((v_j)_{j\not=i})\end{array}\Bigg\}\ \colon\ \begin{array}{l} (e_j)_{j\not=i}\in \prod\limits_{j\not=i}E_j \\ f:\prod\limits_{j\not=i}e_j\to E_i\end{array}\right\}\]
    \end{defn}
    It can be interpreted as the $k-1$ other players each picking an measurement $e_j\in E_j$ and for each product outcome of these the player $i$ picks a measurement in $E_i$.
    \begin{defn}$k$-partite Products\\
        Let $H_i=(V_i,E_i)$ for $i\in[1,k]$ be $k$ hypergraphs. The \textbf{minimal Foulis-Randall product} of the $H_i$, denoted $H_{FR}^{min}=\prescript{min}{1\leq i\leq k}{\bigotimes}H_i$, is defined as: \[H_{FR}^{min}=\Bigg(\prod_{i=1}^kV_i\ ,\ \bigcup\limits_{i=1}^kE_{i\to[1,k]\setminus i}\Bigg)\] The \textbf{complete Foulis-Randall product} of the $H_i$, denoted $\overline{H}_{FR}=\overline{\bigotimes_{1\leq i\leq k}}H_i$, is defined as: \[\bar H_{FR}=\overline{H_{FR}^{min}}\]
    \end{defn}
    
    And with these products comes the generalization of the property of $\mathcal{G}(\prescript{min}{1\leq i\leq k}{\bigotimes}H_i)$: a probabilistic model $p\in\mathcal{G}(H_1\times...\times H_k)$ lies in $\mathcal{G}(\prescript{min}{1\leq i\leq k}{\bigotimes}H_i)$ if and only if it satisfies the $k$-partite no-signaling equations.


    \begin{defn}$k$-partite Game\\
        Let $H=(V,E)=\overline{{\bigotimes}}_{i=1}^kB_i$ with $B_i=(V_i,E_i)\in\mathcal{B},\ \forall i\in[1,k]$.\\
        The edges of $H$ of the form of $\prod_{i=1}^ke_i$ with $e_i\in E_i,\ \forall i\in[1,k]$ are called \textit{questions} of $H$ and the set of questions is denoted by $Q_E$.
        A function $r:Q_E\to\mathcal{P}(V)$ such that $\forall e\in Q_E,\ r(e)\subseteq e$ is called a \textit{rule} on $H$. For any rule $r$, the \textit{winning outcomes} are the vertices in $W_r=\bigcup_{e\in Q_E}r(e)$ and the \textbf{game on $H$ under the rule $r$} is the subsequent hypergraph \[H_r=\Bigg(W_r\ ,\ \{\ e\cap W_r\ \colon\ e\in E\ \}\Bigg)=H_{W_r}\]
    \end{defn}
 
 Note that the translation from compatible measurements to the event based hypergraph model is well known and can be found in the appendix of \cite{acin2015combinatorial}.  A  straightforward representation   of $k-$partite non local games is obtained in the compatible measurements as each player's measurements have to be compatible with all the other player's measurement and each player can chose exactly one measurement. Therefore in the compatibility measurements each player has a set of vertices two by two disjoint and the sets are connectected as complete multipartite graphs.

\section{Contextuality as Bipartite Non-Locality}
    The first result in terms of conditional contextuality is that any scenario can be found in a bipartite game when the two players operate on uncorrelated scenarios as large as the intended contextuality scenario.
    \begin{thm}General Bipartite conditional contextuality\\
    Let $H=(V,E)$ be a hypergraph such that $|E|=m$ and $max_{e\in E} |e| =d$. Then, $H$ has a $2$-partite conditional interpretation as a game on $B_{2,m,d}$.
    \end{thm}
    \begin{proof}
    First, let's formally define the sought game and then construct the conditional interpretation. Consider the two $1$-local scenarios $H_a=(V_a,E_a)$ and $H_b=(V_b,E_b)$ such that \[H_a=H_b=\bigg(\ \big\{v|e\ \colon\ e\in E,\ v\in e\big\}\ ,\ \big\{\{v|e\ \colon\ v\in e\}\ \colon\ e\in E\big\}\ \bigg)\] Those are the factors scenarios. For each $v\in V$, there are, in $V_a$ and $V_b$, as many $clones$ of $v$ as edges of $H$ which contains $v$. Each edge $\{v|e\ \colon\ v\in e\}$ in $E_a$ and $E_b$ can be interpreted as a decorrelated version of $e$. $H_a$ and $H_b$ are thus two uncorrelated scenarios, observe that there is already a labeling induced by the names of the vertices. In the product hypergraph $H_a\otimes H_b$, the edge $\{vv'|ee'\ \colon\ v\in e,\ v'\in e'\}$ is called the $question$ on the edges $(e,e')$ and will be denoted by $ee'$ . Finally, the sought game $G_{ab}=(V_{ab},E_{ab})$ is the game on $H_a\otimes H_b$ under the standard rule $r$, $G_{ab}=(H_a\otimes H_b)_r$ such that \[\forall e,e'\in E,\ r(ee')=\big\{vv'|ee'\ \colon\ v=v'\ or\ \{v,v'\}\cap e\cap e'=\emptyset\big\}\] Now let's prove that $H$ is a conditional scenario of $G_{ab}$. To show that, consider for each $v\in V$ an edge $e^v\in E$ such that $v\in e^v$, this creates an injection $\phi:v\in V\mapsto vv|e^ve^v\in V_{ab}$. The (stronger) result which will be proved is that any choice of the edges $e^v$ makes a valid labeling of $H$.
    \begin{itemize}
        \item Let $p_{ab}\in\mathcal{G}(G_{ab})$ be a probabilistic model and $p=p_{ab}\circ\phi$. In order to show $p\in\mathcal{G}(H)$, i.e. \[\forall e\in E,\ \sum_{v\in e}p(v)=\sum_{v\in e}p_{ab}(vv|e^ve^v)=1,\] consider an edge $e=\{v_1,...,v_k\}\in E$. Let's show that \[\forall i\in[1,k],\ p_{ab}(v_iv_i|e^{v_i}e^{v_i})=p_{ab}(v_iv_i|ee)\] and then use the question on $(e,e)$ to get the result.\\
        Consider for each $i\in[1,k]$ the two following joint measurement edges of $H_a\otimes H_b$, respectively in $E_{a\to b}$ and $E_{a\gets b}$:
        \begin{align*}
            (e_{a\to b})_i=\big(\{v_i|e\}\times e\big)\ &\cup\ \bigcup_{j\not=i}\ \{v_j|e\}\times e^{v_i}\\
            (e_{a\gets b})_i=\big(e^{v_i}\times\{v_i|e^{v_i}\}\big)\ &\cup\bigcup_{v'\in e^{v_i}\setminus\{v_i\}}e\times \{v'|e^{v_i}\}
        \end{align*}
        Now, in the game $G_{ab}$ these edges lost vertices thanks to the rule $r$ applied to the questions on $(e,e)$, $(e,e^{v_i})$ and $(e^{v_i},e^{v_i})$. In $E_{ab}$, the edge $(e_{a\to b})_i$ was restricted to:
        \[\{v_iv_i|ee\}\ \cup\bigcup_{\substack{j\not=i\\v_j\in e^{v_i}}}\{v_jv_j|ee^{v_i}\}\ \cup\bigcup_{\substack{j\not=i\\v_j\not\in e^{v_i}}}\{v_jv'|ee^{v_i}\ \colon\ v'\in e^{v_i}\setminus e\}\]
        Then, observe that these two unions can be rewritten as:
        \begin{align*}
            \bigcup_{\substack{j\not=i\\v_j\in e^{v_i}}}\{v_jv_j|ee^{v_i}\}\ =\{ww|ee^{v_i}\ &\colon\ w\in (e\cap e^{v_i})\setminus\{v_i\}\}=\bigcup_{\substack{v'\in e^{v_i}\setminus\{v_i\}\\v'\in e}}\{v'v'|ee^{v_i}\}\\
            \bigcup_{\substack{j\not=i\\v_j\not\in e^{v_i}}}\{v_jv'|ee^{v_i}\ \colon\ v'\in e^{v_i}\setminus e\}&=\{v_jv'|ee^{v_i}\ \colon\ v_j\in e\setminus e^{v_i},\ v'\in e^{v_i}\setminus e\}\\&=\bigcup_{\substack{v'\in e^{v_i}\setminus\{v_i\}\\v'\not\in e}}\{v_jv'|ee^{v_i}\ \colon\ v_j\in e\setminus e^{v_i}\}
        \end{align*}
        And $(e_{a\gets b})_i$ was restricted to:
        \[\{v_iv_i|e^{v_i}e^{v_i}\}\ \cup\bigcup_{\substack{v'\in e^{v_i}\setminus\{v_i\}\\v'\in e}}\{v'v'|ee^{v_i}\}\ \cup\bigcup_{\substack{v'\in e^{v_i}\setminus\{v_i\}\\v'\not\in e}}\{v_jv'|ee^{v_i}\ \colon\ v_j\in e\setminus e^{v_i}\}\]
        Thus, these two unions form a subset $W\subseteq V_{ab}$ such that $\{v_iv_i|ee\}\ \cup\ W$ and $\{v_iv_i|e^{v_i}e^{v_i}\}\ \cup\ W$ are both edges in $E_{ab}$, i.e.
        \[p_{ab}(v_iv_i|ee)=1-p_{ab}(W)=p_{ab}(v_iv_i|e^{v_i}e^{v_i})\]
        Hence, $\forall i\in[1,k],\ p_{ab}(v_iv_i|ee)=p_{ab}(v_iv_i|e^{v_i}e^{v_i})$ and so: \[\sum_{v_i\in e}p(v_i)=\sum_{v_i\in e}p_{ab}(v_iv_i|e^{v_i}e^{v_i})=\sum_{v_i\in e}p_{ab}(v_iv_i|ee)\] Finally, $\{v_iv_i|ee\ \colon\ v_i\in e\}\in E_{ab}$, therefore $\sum\limits_{v_i\in e}p(v_i)=1$.\\
        \begin{figure}[ht]
            \centering
            \includegraphics[width=13.5cm]{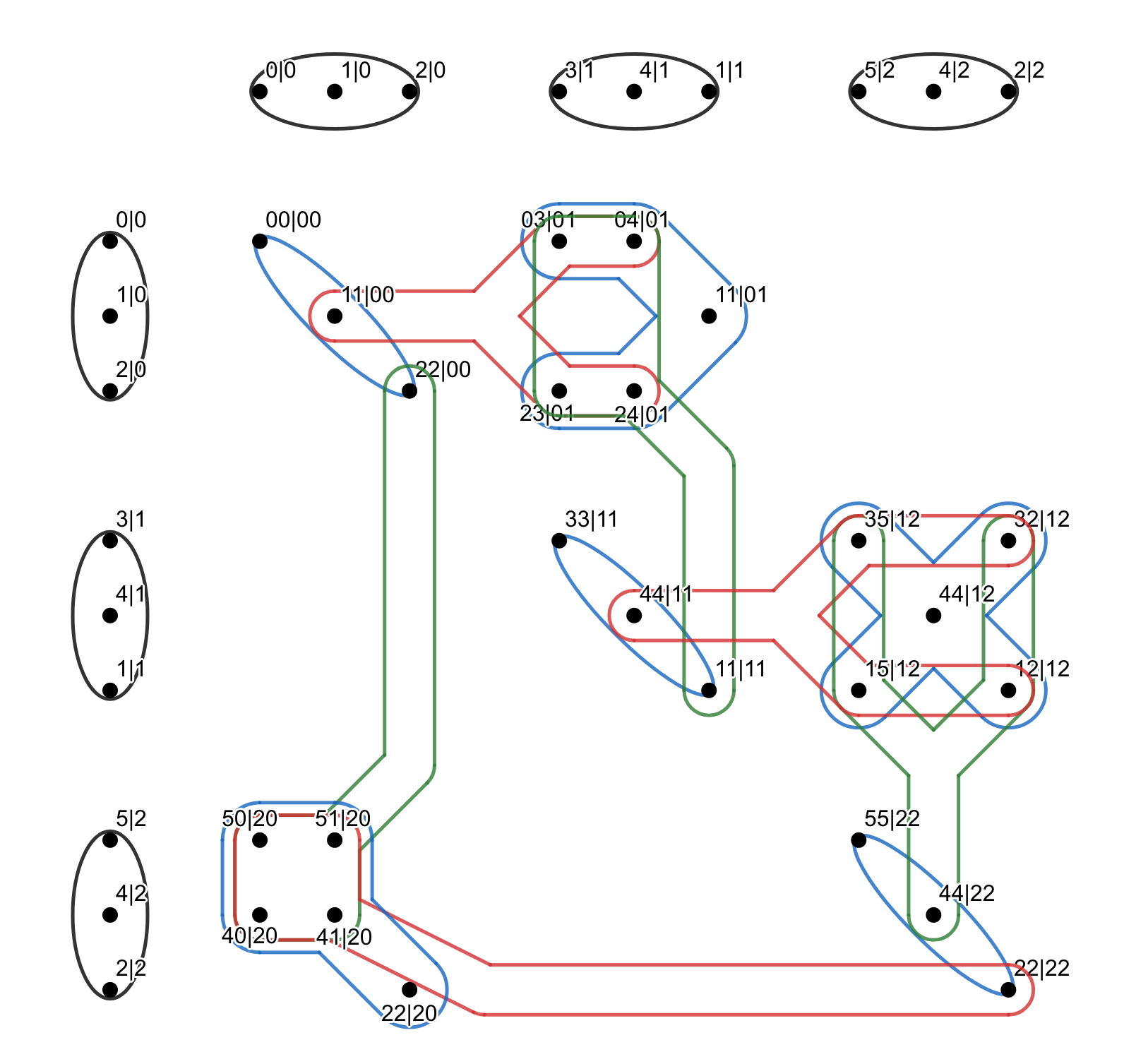}
            \caption{Example of the construction for the $\Delta_3$ scenario (cf. Figure \ref{fig:delta}). The rules are in blue, the edges used in this proof are in red and green (also half of the vertices are omitted because the construction is symmetrical). The vertices labeled  $00|00, 11|00, 22|00, 33|11, 44|11$ and $55|22$ have valid probability distributions in bijection with the vertices of $\Delta_3$. The first three represent one edge, the second diagonal edge uses a clone of $11||00$  that is  $11|11$ and the third diagonal edge uses two clones.}
            \label{fig:proof1}
        \end{figure}
        \item Let $p\in\mathcal{G}(H)$ be a probabilistic model, and let's define a candidate for its extension to $\mathcal{G}(G_{ab})$ as $p_{ab}:V_{ab}\to[0,1]$ such that \[\forall e,e'\in E,\ p_{ab}(vv'|ee')=\begin{dcases}p(v)&\text{if }v=v'\\\frac{p(v)p(v')}{1-p(e\cap e')}&\text{if }\{v,v'\}\cap e\cap e'=\emptyset\text{ and }p(e\cap e')<1\\0&\text{if }\{v,v'\}\cap e\cap e'=\emptyset\text{ and }p(e\cap e')=1\end{dcases}\] Then, it follows that $\forall v\in V,\ (p_{ab}\circ\phi)(v)=p_{ab}(vv|e^ve^v)=p(v)$, hence $p_{ab}\circ\phi=p$. Now, in order to show $p_{ab}\in\mathcal{G}(G_{ab})$, i.e. \[\forall e\in E_{ab},\ \sum_{v\in e}p_{ab}(v)=1,\] consider $e_{ab}\in E_{ab}$. As $G_{ab}$ was obtained from the product $H_a\otimes H_b$, there exists a joint measurement edge $e_{jm}\in E_{a\to b}\cup E_{a\gets b}$ such that $e_{ab}=e_{jm}\cap V_{ab}$. Let's consider the case where $e_{jm}\in E_{a\to b}$, the other is treated symmetrically. By definition of $E_{a\to b}$, there exists the edges $e\in E,\ e_a=\{v|e\ \colon\ v\in e\}\in E_a$ and the function $f:e_a\to E_b$ such that $e_{jm}=\bigcup\limits_{v_a\in e_a}\big(\{v_a\}\times f(v_a)\big)$. Therefore:
        \begin{align*}
        \sum_{u\in e_{ab}}p_{ab}(u)&=\sum_{u\in e_{jm}\cap V_{ab}}p_{ab}(u)=\sum_{v_a\in e_a}\ \sum_{u\in(\{v_a\}\times f(v_a))\cap V_{ab}}p_{ab}(u)\\&=\sum_{v\in e}\ \sum_{u\in(\{v|e\}\times f(v|e))\cap V_{ab}}p_{ab}(u)\\&=\sum_{v\in e}\ \sum_{\substack{v'|e'\in f(v|e)\\v=v'\ or\ \{v,v'\}\cap e\cap e'=\emptyset}}p_{ab}(vv'|ee')
        \end{align*}
        For each vertex $v\in e$ there is $f(v|e)\in E_b$, so consider the edge $e'\in E$ such that $f(v|e)=\{v'|e'\ \colon\ v'\in e'\}$.\\If $v\in e'$ then $\forall v'\in e',\ v\in \{v,v'\}\cap e\cap e'$ thus: \[\sum_{\substack{v'|e'\in f(v|e)\\v=v'\ or\ \{v,v'\}\cap e\cap e'=\emptyset}}p_{ab}(vv'|ee')=p_{ab}(vv|ee')=p(v)\] and if $v\not\in e'$ then:
        \begin{align*}
        \sum_{\substack{v'|e'\in f(v|e)\\v=v'\ or\ \{v,v'\}\cap e\cap e'=\emptyset}}p_{ab}(vv'|ee')&=\sum_{v'\in e'\setminus e}\frac{p(v)p(v')}{1-p(e\cap e')}=\frac{p(v)p(e'\setminus e)}{1-p(e\cap e')}\\&=\frac{p(v)p(e'\setminus e\cap e')}{1-p(e\cap e')}=\frac{p(v)(p(e')-p(e\cap e'))}{1-p(e\cap e')}\\&=\frac{p(v)(1-p(e\cap e'))}{1-p(e\cap e')}=p(v)
        \end{align*}
        Hence, the result: \[\sum_{u\in e_{ab}}p_{ab}(u)=\sum_{v\in e}p(v)=1\] 
    \end{itemize}
    \end{proof}
    In that respect any contextuality scenario is captured by the joint experiment of two players who operate on a hypergraph larger than the scenario. The size of the non-local product which captures it  using the construction of Theorem 4.1 is   the square of its size.\par
\begin{figure}[ht!]
        \centering
        \includegraphics[width=13cm]{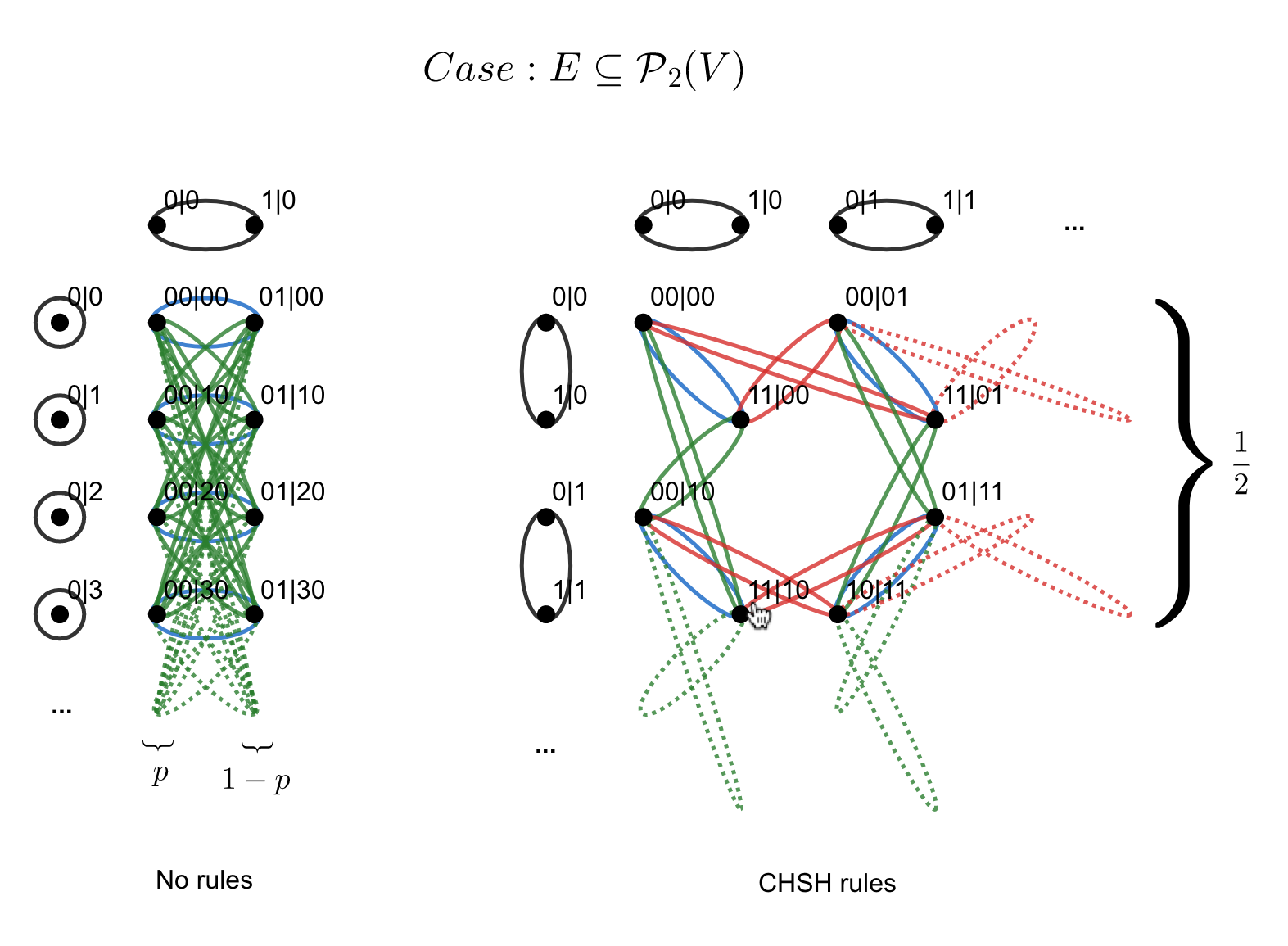}
        \caption{If the graph is bipartite (no odd-length cycles) then the left construction shows that it's in a product $B_{1,1,2}\oplus B_{1,m,1}$, otherwise it's in a CHSH-like scenario where you extend the factors by adding $B_{1,2,2}$}
        \label{fig:graph}
    \end{figure}

    This result can be improved, for example in Figure $\ref{fig:graph}$ it is clear that any connected scenario with only binary edges (a graph seen as a hypergraph) has a $2$-partite conditional interpretation with linear-size factors with respect to $|V|$.\par
    \begin{thm}Bipartite Conditional Contextuality of Graphs\\
        Let $H=(V,E)$ be a connected hypergraph such that its edges are binary, i.e. $\forall e\in E,\ |e|=2$. Then, $H$ has a $2$-partite conditional interpretation as a game with a size of $O(|V|)$.
    \end{thm}
\begin{proof}
    There are two types of connected graphs: bipartite graphs and graphs which contain an odd-length cycle. In the first case, consider $U_1\uplus U_2$ the partition of $V$ such that $\forall e\in E,\ \exists u_1\in U_1,\ u_2\in U_2,\ e=\{u_1,u_2\}$ and $|U_i|=m_i$. The only models of $H$ are of the form $p(u_1)=q,\ p(u_2)=1-q$ with $q\in[0,1]$ because $H$ is bipartite and connected. The product $B_{1,max(m_1,m_2),1}\oplus B_{1,1,2}$ has a canonical labelling of the form $0b|x0$ with $b\in\{0,1\},\ x\in[0,max(m_1,m_2)-1]$. The conditional interpretation of $H$ comes by assigning the vertices of $U_i$ to the the product vertices with labels $0(i-1)|x0$ because the only models of this product are of the form $p(00|x0)=q,\ p(01|x0)=1-q$ with $q\in[0,1]$. In the second case, consider product $B_{1,2m,2}\oplus B_{1,2n,2}$ such that $8mn\geq|V|$, its canonical labeling of the form $ab|xy$ with $a,b\in\{0,1\},\ x\in[0,m-1],\ y\in[0,n-1]$, and the game on this product with an extension of the CHSH rules: $r(xy)=\{ab|xy\colon a=b\iff(x\equiv0[2])\vee(y\equiv0[2])\}$. This game is $mn$ games of CHSH so this gives a scenario with $8mn$ vertices and only one model which assigns $1/2$ to all of them. As $H$ is connected and contains an odd-length cycle it has only one model which assigns $1/2$ to all the vertices hence the conditional interpretation.
    \end{proof}

\section*{Conclusion}
 We have shown that any contextuality scenario can be captured by a bipartite non local game using the notion of conditional contextuality, and that the construction we provide is not optimal in size. It would be interesting to characterise the subset that could be perfectly captured by non-locality.  In addition from the foundational question about how general each setting is, it might have some potential applications offering a new point of view to test or certify   contextual scenarios using two-player games.

A natural link to be investigate is the relation between this "conditional contextuality" and almost quantum correlations analysed in \cite{gonda2018almost,Sainz_2018} and also with  the notion  of "no detection events" used recently by Kunjwal \cite{Kunjwal2019beyondcabello}  to relate graph contextuality of \cite{cabello2010non} to hypergraph contextuality.

 Another intriguing and interesting direction for further analysis is the multipartite case: if we consider players with binary  inputs/outputs,
 is it possible  to capture  any contextual scenario on $n$ vertices as  a $k$-player game, and how does the optimal number of player relate to other measures of multipartiteness.

\section*{Acknowledgements}
The authors would like to thank Elie Wolfe for very useful comments and feed-back.
This research was supported through the program \emph{``Investissements d'avenir''} (ANR-15-IDEX-02) program of
the French National Research Agency.

\section*{Appendix }
    In order to prove the consistency of VCZ equivalence - i.e. for all contextuality scenarios $H_1$ and $H_2$ such that $H_1\sim_{vcz}H_2$ (any choice of) $H_1^{vcz}$ and $H_2^{vcz}$ are observationally equivalent- we will need to use the following lemmas.
    \begin{lema}[Structure of $V^{vcz}$]    
        Let $H=(V,E)$ be a hypergraph. A \textit{maximal set of indistinguishable non zero weighted vertices} is a subset $W\subseteq V$ such that $W$ doesn't contain zero weighted vertices, it can be contracted in $H$ and it is a maximal element of $\mathcal{P}(V)$ satisfying these properties, i.e.:
        \begin{align}
            \forall w\in W,\ \exists p\in\mathcal{G}(H),\ p(w)>0&\\
            \forall e\in E,\ (W\subseteq e)\vee(W\cap e=\emptyset&)\\
            (W'\supseteq W)\wedge(W'\text{ satisfies (1) and (2)})\Rightarrow&\ W'=W
        \end{align}
        \sloppy
        Then, any choice of $H^{vcz}=(V^{vcz},E^{vcz})$ satisfies that $V^{vcz}$ is isomorphic to $\mathcal{M}(V)=\{\text{maximal sets of indistinguishable non zero weighted vertices}\}$.
    \end{lema}
\begin{proof}    A choice of $V^{vcz}$ is done by removing all zero weighted vertices and then contracting the remaining vertices so that $|V^{vcz}|$ is minimal. Consider $V^*\subseteq V$ the set of non zero weighted vertices of $V$, now $\mathcal{M}(V)$ is a partition of $V^*$ and so choosing a $V^{vcz}$ is just picking exactly one vertex in each element of $\mathcal{M}(V)$, hence the isomorphism.
\end{proof} 
    \begin{lema}[Independence of zero-reduction]
        Let $H=(V,E)$ be a hypergraph such that $V$ contains a zero weighted vertex $v_0$ and consider the induced sub-hypergraph $H_{V\setminus\{v_0\}}$. Then any choice of $H^{vcz}$ and $H_{V\setminus\{v_0\}}^{vcz}$ are observationally equivalent.
    \end{lema}
    \begin{proof}
    Let $H^{vcz}=(V^{vcz},E^{vcz})$ and $H_{V\setminus\{v_0\}}^{vcz}=(V'^{vcz},E'^{vcz})$ be the choices. Since $v_0$ is zero weighted $\mathcal{M}(V)=\mathcal{M}(V\setminus\{v_0\})$, so both $V^{vcz}$ and $V'^{vcz}$ are isomorphic to $\mathcal{M}(V)$ thus the two are isomorphic. Hence their completion is the same up to this isomorphism and so $H^{vcz}$ and $H'^{vcz}$ are observationally equivalent because removing edges by virtual equivalence preserve the set of probabilistic models.
    \end{proof}
    \begin{lema}[Independence of contraction]
        Let $H=(V,E)$ be a hypergraph, $W\subseteq V$ such that $W$ can be contracted in $H$ and consider the contraction $H_{V\setminus(W\setminus W')}$ of $H$ with $W'\subseteq W,\ W'\not=\emptyset$. Then any choice of $H^{vcz}$ and $H_{V\setminus(W\setminus W')}^{vcz}$ are observationally equivalent.
    \end{lema}
    \begin{proof}
    Let $H^{vcz}=(V^{vcz},E^{vcz})$ and $H_{V\setminus(W\setminus W')}^{vcz}=(V'^{vcz},E'^{vcz})$ be the choices. Since $W'\subseteq W,\ W'\not=\emptyset$ $\mathcal{M}(V)$ and $\mathcal{M}(V\setminus(W\setminus W'))$ only differ on one element (in the case $W$ contains a non zero weighted vertex, if not there are identical): there exists $U\subseteq V\setminus W$ such that $\mathcal{M}(V)\setminus\{W\cup U\}=\mathcal{M}(V\setminus(W\setminus W'))\setminus\{W'\cup U\}$. So $\mathcal{M}(V)$ is isomorphic to $\mathcal{M}(V\setminus(W\setminus W'))$ by assigning $W\cup U$ to $W'\cup U$ and thus $V^{vcz}$ and $V'^{vcz}$ are isomorphic. Hence their completion is the same up to this isomorphism and so $H^{vcz}$ and $H'^{vcz}$ are observationally equivalent because removing edges by virtual equivalence preserve the set of probabilistic models.
    \end{proof}
    \begin{lema}[Independence of virtual inclusion]
        Let $H=(V,E)$ be a hypergraph such that $e\subseteq V$ is a virtual edge of $H$ and consider the hypergraph $H'=(V,E\cup\{e\})$. Then any choice of $H^{vcz}$ and $(H')^{vcz}$ are observationally equivalent.
    \end{lema}
   \begin{proof}
    The zero weighted vertices of $H$ and $H'$ are the same because they have the same probabilistic models. They also have the same maximal sets of indistinguishable non zero weighted vertices because the virtual edge $e$ can't "split" such a set: if $e'$ contains only a subset of $W\in\mathcal{M}(V)$ we can find two models of $H$ which have a different value of $e'\cap W$ but the same values on the other vertices of $e'$ because the vertices of $W$ can be interchanged. So any choice of $H^{vcz}$ and $(H')^{vcz}$ gives $V^{vcz}$ isomorphic to $(V')^{vcz}$, hence $H^{vcz}$ and $(H')^{vcz}$ are observationally equivalentby the same previous reasoning.
            \end{proof}

    Theorem \ref{thm:vcz} is then proved by considering that two scenarios $H_1$ and $H_2$ such that $H_1\sim_{vcz}H_2$ can be obtained from one another step by step with zero-reduction, contraction and virtual inclusion. So this creates a chain of observationally equivalent scenarios and thus any choice for $H_1^{vcz}$ and $H_2^{vcz}$ are observationally equivalent.

\bibliographystyle{plain}

\end{document}